\documentclass[12pt]{iopart}
\usepackage{graphicx}

\begin{document}

\title{Stoke's efficiency and it's stochastic properties}

\author{Mamata Sahoo$^{1}$, A. M Jayannavar$^2$}
\address{
$^1$ Computational Modelling $\&$ Simulation Section, National Institute for Interdisciplinary Science and Technology, Thiruvananthapuram -695019, India\\
$^2$ Institute of Physics, Sachivalaya Marg, Bhubaneswar, Orissa, India}

\ead{mamata.sahoo@niist.res.in, jayan@iopb.res.in}
\begin{abstract}                          
We study the Stoke's efficiency and it's fluctuating properties in the case of a spatial asymmetric ratchet potential with a temporal asymmetric driving force from adiabatic to nonadiabatic regime. Our numerical investigations show that the average Stoke's efficiency and the average current decrease with the frequency of driving. For low frequency of driving, i.e., in the case of an adiabatic regime, we reproduced the analytical results supporting our numerical simulations. By evaluating the probability distribution, $p(\eta_{s})$ for Stoke's efficiency, $\eta_{s}$ we focus on the stochastic properties of Stokes efficiency. We find that in most of the parameter space, fluctuations in $\eta_{s}$ are comparable to or larger than the mean values. In such a situation one has to study the full probability distribution of $\eta_{s}$. With increase in frequency of driving, the distribution becomes multipeaked. At the same time the average Stoke's efficiency decreases with increase in frequency of drive. For high frequency of driving, the distribution develops a peak across zero. Further increase in frequency this peak gets sharper. And finally at sufficiently high frequency we get a strong peak across zero indicating that there is no effective transport in this regime. 
\vskip.5cm
%{\small{\textit{PACS :} 05.40.-a; 05.70.lw}} 
%{\small{\textit{Keywords:}Ratchets, entropy production, noise, energy loss.}}
\end{abstract}
%\end{frontmatter}


\noindent{\it keywords}:Ratchets, Efficiency, Fluctuations

\maketitle

\section{Introduction}
Over the last several decades the study of interplay of noise and nonlinearity has become a major subject of research in various multidisciplinary areas\cite{Julicher1,Astumian,Reimann,jayan, Gammaitoni, Hangi}. These studies include the noise induced directed transport, stochastic resonance, noise induced stability of unstable states, noise induced phase transition, noise induced ordering etc. In most of the above mentioned cases the presence of noise is very much essential. In the literature, the noise induced directed transport in periodic extended structures in the absence of overall net bias has been extensively studied. Two basic ingredients are required for the posilibity of such directed transport, namely, the system should be driven out of equilibrium and there should be some asymmetry (either temporal or spatial) along with nonlinearity in the system. The systems consisting of Brownian particles and operating with the minimal conditions for such directed transport are known as Brownian ratchets/Brownian motors \cite{marches, gommers}. There is increasing interest from physicists, biologists and engineers on the study of so called ratchets or Brownian motors \cite{Ajdari1, Prost1, Linke, Astumian1}. Based on the various ways of introducing the asymmetry or nonlinearity in the systems, there are various kinds of Brownian ratchet models like flashing ratchets, rocking ratchets, time-asymmetric ratchets and inhomogeneous ratchets\cite{Reimann}.

The main motivation in the study of performance characteristics of Brownian ratchet is the notion of efficiency of energy transduction from the thermal fluctuations \cite{linke1}.  With the help of stochastic energetic  formalism \cite{parrondo, sekimoto, sekimoto1} proposed by Sekimoto, efficiency can be defined in a wide class of ratchet models \cite{kamgawa, Dan}. Different kinds of efficiency (thermodynamic, Stoke's and generalized) for ratchet models have been studied in detail both analytically and numerically. Since the Brownian ratchet operates out of equilibrium, there is always an unavoidable heat transfer to the medium/environment, which makes it to be less efficient. It has been noticed that Brownian motors based on the principle of flashing ratchet models result in low efficiency. On the other hand, Brownian motors based on the adiabatically changing potentials, e.g., rocked ratchet models exhibit high value in efficiency \cite{parrando1, sokolov, hernandez,jstat, mamata1}. In this work, we are mainly interested in the case of spatial asymmetric rocked ratchet models.  

In all kinds of ratchet models, it is to be emphasized that the Brownian particle moves in a periodic potential system and hence it ends up with the same potential energy even after crossing over to the adjacent potential minimum. Hence there is no extra energy stored in the particle which can be used for a given purpose. In order to extract work/energy out of it's motion, it is necessary to apply an external load force against which the particle moves and stores energy in the form of potential energy \cite{parrondo,parrando1,kamgawa-parrondo}. In this context the efficiency of the ratchet can be defined as the output work against the load force per input energy given to the system. This definition of efficiency is called as thermodynamic efficiency and it's value is zero when there is no applied load in the ratchet \cite{parrondo}. The ratchets are not designed to lift the load all the time. Alternatively, there are various ways of defining the efficiency in the ratchet based on the purpose of the work given to the ratchet. In this work, we are mainly interested in the notion of Stoke's efficiency, which is defined as the ratio of the rate of work done against the mean viscous drag and the rate of input energy given to the ratchet \cite {munakata, oster, wang}. 

The Brownian ratchets operate at microscopic length scale. For the systems operating at this length scale, the energy exchanged (i.e., in the form of mechanical energy, chemical energy etc.) between the system and it's environment is of the order of thermal energy (few $k_{B} T$, $k_{B}$ being the Boltzmann constant and $T$ being the temperature). Thermal fluctuations play a predominant role, thereby exhibiting distinctly different behavior from that of macro systems \cite{Bustamante}. In the present work, we are mainly interested in the study of fluctuating properties of Stoke's efficiency in case of a spatial and temporal asymmetric rocked ratchet model. Recently the enhancement of Stoke's efficiency by noise was studied in a microscale domain \cite{luczka}. In this study, we are mainly interested on the stochastic properties of Stoke's efficiency and it's probability distribution from adiabtic to nonadiabatic limit. We show that in most of the parameter space, fluctuations dominate the mean values. As we increase the frequency of the drive the mean Stoke's efficiency and the average current decreases. 

\begin{figure}[hbp!]
\begin{center}
\input{epsf}	
\includegraphics [width=3in,height=2.5in]{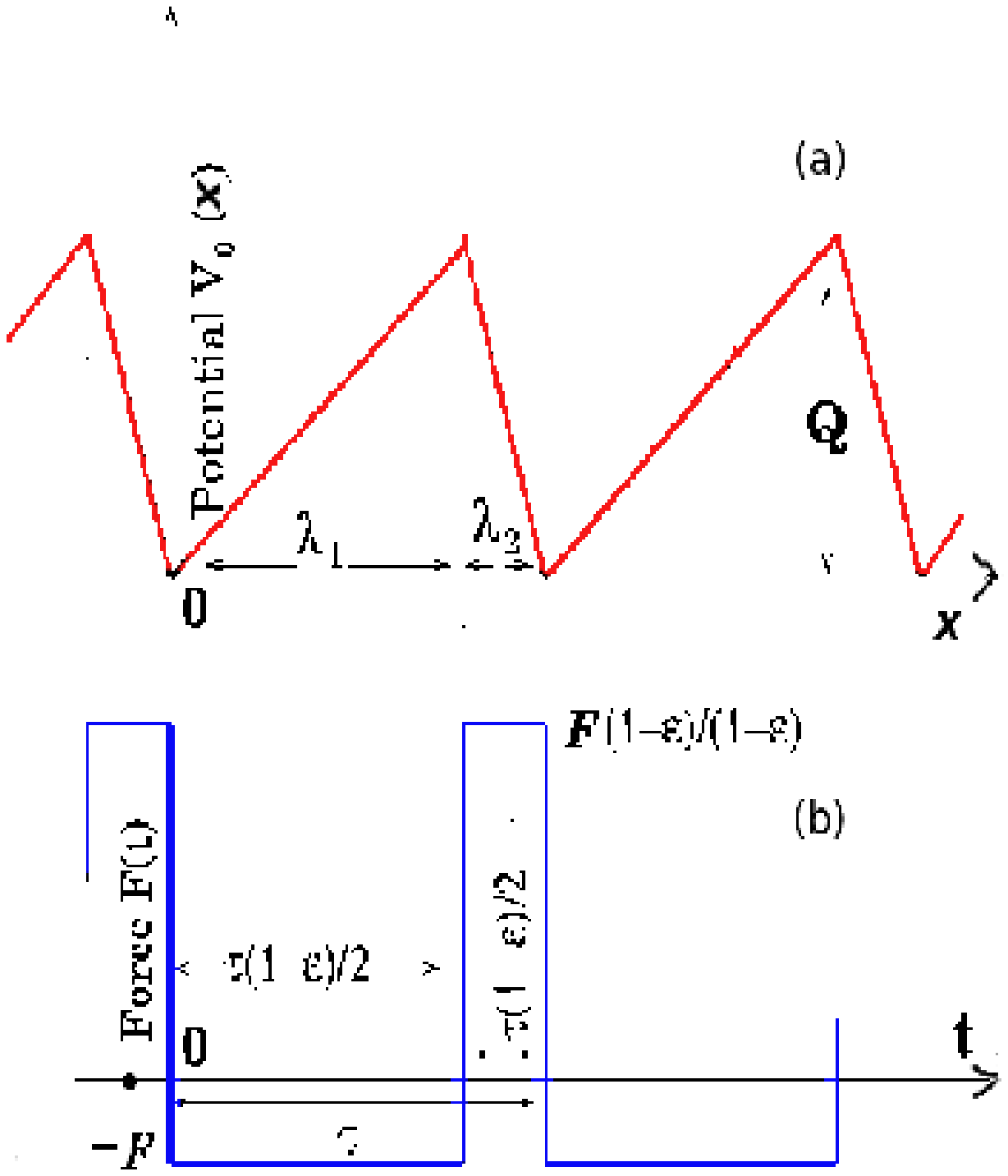}
\caption{Schematic diagram of a spatial asymmetric ratchet potential as a function of coordinate $x$ 
and the external time asymmetric driving force $F(t)$ as a function of $t$. }
\label{sawtooth}
\end{center}
\end{figure}

\section{The Model}
We consider the dynamics of an overdamped Brownian particle in a periodic spatial asymmetric ratchet potential $V(x)$ in the presence of an external time asymmetric driving force $F(t)$. The motion is governed by the overdamped Langevin equation \cite{Dan}
\begin{equation}
 \gamma {\dot{x}} = -V^\prime(x)+F(t)+\xi(t),\label{lang}
 \end{equation}
where $x(t)$ denotes the instantaneous position of the particle and $\gamma$ being the friction coefficient. $\xi(t)$ is the randomly fluctuating Gaussian thermal noise satisfying the properties, 
$<\xi(t)\xi(t^\prime)>\,=\,(2\,{k_BT}/\gamma) \delta(t-t^\prime)$ and $<\xi(t)>$=0. The angular bracket $<...>$ denote the ensemble average over all the realizations of noise. 
A schematic diagram of the ratchet potential $V(x)$ along with the external drive, $F(t)$ is shown in Fig.1.
\begin{eqnarray}
V(x) &=& \frac{Q}{\lambda_1} x, \,\,\,\,\,\,\,\,x\leq \lambda_1 \nonumber \\
 &=& \frac{Q}{\lambda_2} (1-x),  \,\,\lambda_1<x\leq \lambda,
\end{eqnarray}
where $Q$ is the height of the ratchet potential. $\lambda=\lambda_{1}+\lambda_{2}$, is the spatial periodicity of the ratchet potential, which is set to unity. $\Delta=\lambda_{1}-\lambda_{2}$ represents the spatial asymmetric parameter which characterizes the asymmetry of the ratchet potential.\\
  
The temporal asymmetric driving force $F(t)$ is given by
\begin{eqnarray}
F(t)&=& \frac{1+\epsilon}{1-\epsilon}\, F,\,\, (n\tau 
\leq t < n\tau+ \frac{1}{2} \tau (1-\epsilon)), \\ \nonumber
    &=& -F,\,\, (n\tau+\frac{1}{2} \tau(1-\epsilon) < t \leq
(n+1)\tau).\label{ft}
\end{eqnarray}
Here $\epsilon$ represents the temporal asymmetric parameter in the driving force and ranges from $0$ to $1$. $\tau$($=\frac{2\pi}{\omega})$ denotes time period of the external driving force with $\omega$ being the frequency of the drive. $n=0,1,2....$ is an integer.

This model has been studied extensively in the adiabatic limit in \cite{mamata1}. In this reference, the nature of the averaged current and average efficiency is studied in detail. In the adiabatic limit and at temperature, $T \rightarrow 0$ limit, the analytical expression for the time averaged current, $<j>$ is obtained in \cite{jstat}.
\begin{equation}
<j>=j_{1}+j_{2},
\end{equation}
where $j_{1}$ and $j_{2}$ are the fraction of currents in the positive and negative direction over a time period of $\tau(1-\epsilon)/2$ and $\tau(1+\epsilon)/2$ respectively. $j_{1}$ is the current when the external driving force, $F(t)=(\frac{1+\epsilon}{1-\epsilon})F$ and $j_{2}$ is the current when the external driving force, $F(t)=-F$. 

In our present work, we study the unidirectional current as well as the Stoke's efficiency in the adiabatic as well as nonadiabatic regimes. We show that in general the current decreases as we shift from adiabatic to nonadiabatic regime. For the first time we have also studied the nature of the stochastic Stoke's efficiency, $\eta_{s}$ and it's fluctuations. We show that in general average Stoke's efficiency, $<\eta_{s}>$ is not a good physical variable because fluctuations dominate the mean values. Large spatial asymmetry in the ratchet potential decrease the fluctuations in $\eta_{s}$. To this end we have calculated the probability distribution, $P(\eta_{s})$ of stochastic Stoke's efficiency, $\eta_{s}$. Recently stochastic effeciency in heat engines has been studied in great detail in ref. \cite{rana}. The Stoke's efficiency is related to the  rate of work done against the average frictional force to the rate of input energy. Formally stochastic Stoke's efficiency, $\eta_{s}$ at any instant of time can be defined as
\begin{equation}
\eta_{s}=\frac{\gamma \dot{x} <v>}{\dot{E_{in}}}=\frac{\dot{E_{out}}}{\dot{E_{in}}},
\end{equation}

where $\dot{x}$ is the instantaneous velocity and $<v>$ being the average velocity of the Brownian particle. $\dot{E_{in}}=F(t)\dot{x}$ is the rate of input energy. $\dot{E_{out}}$ is the rate of work done against the viscous force, $\gamma <v>$. In this work we are mainly interested in the nature of average Stoke's efficiency, $<\eta_{s}>$ as well as on it's distribution $P(\eta_{s})$. 

Stoke's efficiency $\eta_{s}$, over a cycle $\tau$ is given by

\begin{equation}
\eta_{s}=\frac{\frac{1}{\tau} \int_{0}^{\tau} \gamma \dot{x} <v> dt}{\frac{1}{\tau}\int_{0}^{\tau} F(t) \dot{x} dt}
\end{equation} 

This is a stochastic quantity and for a given period of the drive. It depends on a realization of $\dot{x}$ over the interval. Averaged Stoke's efficiency is now defined as
\begin{equation}
<\eta_{s}>=<\frac{\frac{1}{\tau} \int_{0}^{\tau} \gamma \dot{x} <v> dt}{\frac{1}{\tau}\int_{0}^{\tau} F(t) \dot{x} dt}>
\end{equation}

\section{Numerical Simulation}
We have simulated the overdamped Langevin dynamics using the algorithim of Huens method \cite{Mannela}. From the solution of Langevin equation, we evaluate the averaged current, input energy and Stoke's efficiency. The averaged current is given by
\begin{equation}
<j>=<\frac{x_t-x_0}{t-t_0}>,
\end{equation} 
where $x_t$ represents the instantaneous position of the particle at time $t$ and $x_{0}$ being the position of the particles at time $t_{0}$. We have run the simulation for $10^5 \tau$, $\tau$ being one period of the drive.  The steady state averages were taken just after ignoring the initial transients upto $10^4 \tau$. Each time step is taken to be $0.001$.

\section{Result and Discussion}
This particular ratchet model has been studied theoretically in Ref. \cite{mamata1} in the adiabatic limit. The nature of average Stoke's efficiency, $<\eta_{s}>$ in the adiabatic regime has been studied in detail. In the present work, we are mainly interested on the stochastic properties of Stoke's efficiency, $\eta_{s}$ and hence on its probability distribution, $P(\eta_{s})$ in the adiabatic limit as well as for different frequency of driving, $\omega$. In this particular ratchet model, when the noise strength is very low, i.e., in the deterministic limit there exists potential barriers for the motion of the Brownian particle in both forward as well as backward directions. The potential barrier in the forward direction disappears only when the driving force in the forward direction is larger than the force exerted by the barrier towards the right of the potential, i.e., when $(\frac{1+\epsilon}{1-\epsilon})F>\frac{Q}{\lambda_{1}}$. Similarly the barrier in the backward direction disappears only when $F>\frac{Q}{\lambda_{2}}$. Hence there are critical forces, $F_{c1}=\frac{Q}{\lambda_{1}}(\frac{1-\epsilon}{1+\epsilon})$ and $F_{c2}=\frac{Q}{\lambda_{2}}$ beyond which the barrier disappears in both forward as well as backward direction. It implies that a finite current exists in the ratchet only when the amplitude of driving, $F$ is in between $F_{c1}$ and $F_{c2}$. When barrier disappears in the backward direction, there is a finite current in the backward direction resulting a negative contribution in the net current, $<j>$ in the ratchet. In the deterministic limit, the averaged current, $<j>$ shows a peaking behavior as a function of amplitude of driving force, $F$. The current increases from $F=F_{c1}$ and shows a maximum at around $F=F_{c2}$ and then decreases with $F$. 

We have simulated the averaged current, $<j>$ and the average Stoke's efficiency, $<\eta_{s}>$ as a function of $F$ and plotted in Fig.2 and Fig.3 respectively for different values of temporal asymmetry parameter, $\epsilon$ for a symmetric potential and for temperature $T=0.01$ (close to the deterministic limit). The inset of both the figures shows the same plot with larger temperature $T=0.1$. These results are for $\omega=0.5$ corresponding to the adiabatic limit.  In this limit the forcing $F(t)$ is assumed to change very slowly. It's frequency is smaller than any other frequency related to the relaxation rate in the problem, i.e., the system remains in the steady state at each instant of time. From Fig.2 and Fig.3, we notice that our simulation is in perfect agreement with the analytical obtained in \cite{mamata1}. The Stoke's efficiency follows the nature of average current as expected. In our work we have scaled all the lengths with respect to $\lambda_{1}+\lambda_{2}$ and energies with respect to the height of the ratchet potential, $Q$. All physical quantities are in the dimensionless form. We have set $\gamma=1$.
From the simulation results obtained in Fig.2, we have calculated the values of $F_{c1}$ for three different values of $\epsilon$, i.e., $F_{c1}=1.63$ for $\epsilon=0.1$, $F_{c1}=0.857$ for $\epsilon=0.4$ and $F_{c1}=0.105$ for $\epsilon=0.9$ respectively in the deterministic limit ($T=0.01$). $F_{c2}=2$ remains same for all $\epsilon$ values since $F_{c2}=\frac{Q}{\lambda_{2}}$ is independent of $\epsilon$. These values are in exact agreement with analytically obtained values in \cite{mamata1}. 

\begin{figure}[hbp!]
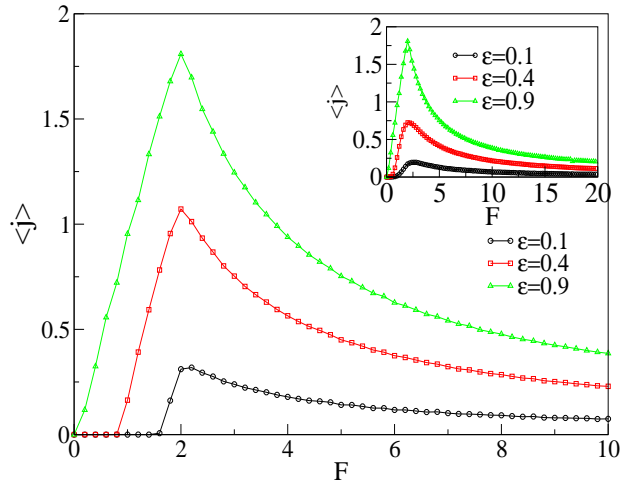

\begin{center}
\input{epsf}
\includegraphics [width=3.2in,height=2.5in]{fig2.eps}
\caption{Average current, $<j>$ as a function of amplitude of driving, $F$ for a symmetric potential $\Delta=0.0$ for different values of $\epsilon$. All other fixed parameters are temperature $T=0.01$, frequency of driving $\omega=0.5$. The inset shows the same plot for temp $T=0.1$.}
\label{sawtooth}
\end{center}
\end{figure}

%\vskip
\begin{figure}
%[hbp!]
\begin{center}
\includegraphics [width=3.2in,height=2.5in]{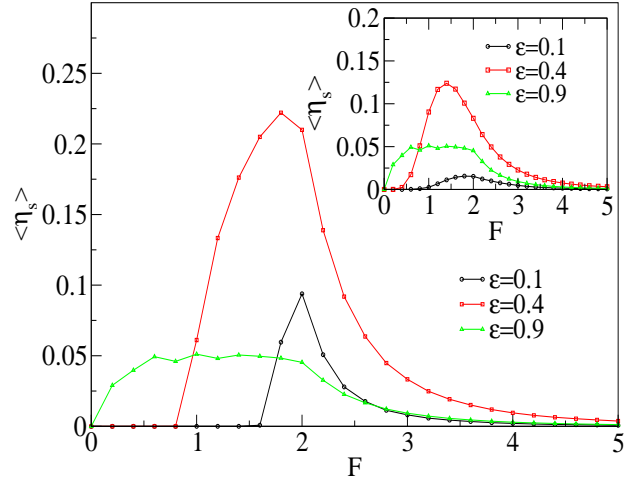}
\caption{The mean Stoke's efficiency, $<\eta_{s}>$ as a function of amplitude of driving, $F$ for a symmetric potential $\Delta=0.0$ for different values of $\epsilon$. All other fixed parameters are temperature $T=0.01$, frequency of driving $\omega=0.5$. The inset shows the same plot for temp $T=0.1$. }
\label{sawtooth}
\end{center}
\end{figure}

Fig.4(a) shows the average current, $<j>$ as  a function of amplitude of driving, $F$. As anticipated earlier in the deterministic limit, current exhibits a peaking behaviour at $F=\frac{Q}{\lambda_{2}}(\simeq 2)$. The average Stoke's efficiency as a function of $F$ follows the nature of currents shown in Fig.4(b). In Fig.4(c), we have plotted the relative variance (fluctuations) of stochastic Stoke's efficiency, $R_{\eta_{s}}=\frac{\sigma_{s}}{<\eta_{s}>}=\sqrt{\frac{(<\eta_{s}^{2}>-<\eta_{s}>^{2})}{<\eta_{s}>^{2}}}$ as a function of $F$. $R_{\eta_{s}}$ shows a decreasing behaviour with $F$ and shows a minimum. However, the minimum is larger than $1$, indicating that the fluctuations, $\sigma_{s}$ in the Stoke's efficiency always dominate their mean values, $<\eta_{s}>$. Hence the average Stoke's efficiency is not a good physical variable in this parameter regime. The distribution of stochastic Stoke's efficiency is studied in Fig.4(d) for various values of amplitude of driving, $F$. For lower $F$ values ($F<F_{c1}$), the distribution shows a single peak. This is because of the presence of potential barrier in both forward as well as backward directions. As a result the particle makes random movement across any one minimum of the ratchet potential. With further increase in $F(>F_{c1})$ values, the barriers in the forward direction disappears and the probability of moving the particle towards right increases. The distribution exhibits multipeaked structure and the mean value shifts towards right indicating that the average Stoke's efficiency, $<\eta_{s}>$ increases with $F$. The multipeaked structure in the distribution is related to the distribution of local average velocities (which we have checked separately in Fig.5) and is due to the presence of barriers. The distribution for negative values of $\eta_{s}$ has a finite weight. This results from atypical realizations of input energy (Fig.6) which show negative tails \cite{sahoo,Jop, sahoo1}. For larger $F$, the distribution becomes narrow and the variance decreases. At the same time the mean value shifts towards left. In this regime, the barriers in both forward as well as backward directions disappear and the particle follows straight paths in both the directions. 

\begin{figure}
%[hbp!]
\begin{center}
\includegraphics [width=4in,height=3.5in]{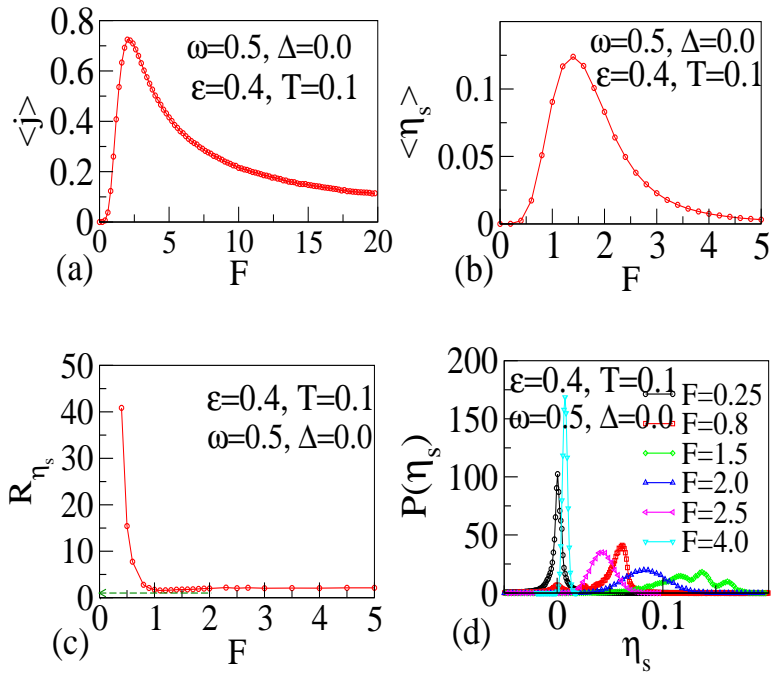}
\caption{$<j>$ as a function of $F$ in (a), $<\eta_{s}>$ as a function of $F$ in (b), relative fluctuation $R_{\eta_{s}}$ as a function of F in (c) and the probability distribution, $P(\eta_{s})$ for different $F$ values in (d). The other fixed parameters are $\omega=0.5$, $\Delta=0.0$, $\epsilon=0.4$ and $T=0.1$ respectively.}
\label{sawtooth}
\end{center}
\end{figure}

\begin{figure}
%[hbp!]
\begin{center}
\includegraphics [width=3.5in,height=3in]{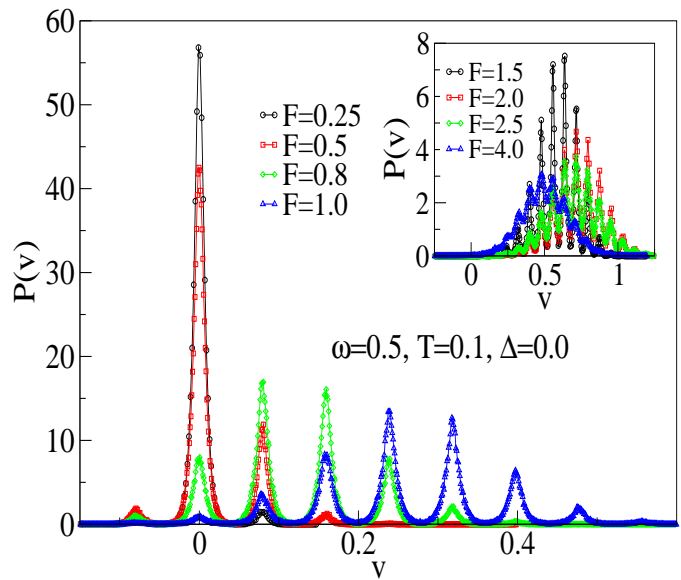}
\caption{The probability distribution of local velocity $v$, $P(v)$ for different $F$ values. The other fixed parameters are $\omega=0.5$, $\Delta=0.0$, $\epsilon=0.4$ and $T=0.1$ respectively.}
\label{sawtooth}
\end{center}
\end{figure}

Next in Fig.7, we have studied the average current, $<j>$, average Stoke's efficiency, $<\eta_{s}>$, Relative fluctuations, $R_{\eta_{s}}$ in $\eta_{s}$ and the probability distribution, $P(\eta_{s})$ with temperature for a symmetric potential in (a), (b), (c) and (d) respectively. We have fixed the other parameters as $F=1$ and $\omega=0.5$. In this parameter regime, in order to have finite current in the ratchet, the minimum value of $\epsilon$ should be greater than $\frac{Q-\lambda_{1}F}{Q+\lambda_{1}F}=0.33$. Hence we have kept fixed the value of $\epsilon$ as $0.4$. Here also we observe that the average current, $<j>$ and average Stoke's efficiency $<\eta_{s}>$ show peaking behaviour with temperature, $T$. The Stoke's efficiency $<\eta_{s}>$ follows the same behaviour of average current $<j>$. However, from Fig.7(a) and 7(b), we notice that there exists a finite current and finite Stoke's efficiency even at zero temperature. This is because of the absence of potential barrier in the forward direction even at zero temperature. In this fixed parameter regime, the crictical value of $F$, $F_{c1}=\frac{Q(1-\epsilon)}{\lambda_{1}(1+\epsilon)}=0.857$ and hence $F>F_{c1}$. Therefore the barrier disappears in the forward direction, as a result we get a finite current and finite Stoke's efficiency even at $T \rightarrow 0$ limit. The relative fluctuations in Stoke's efficiency $R_{\eta_{s}}$ with temperature is studied in Fig.7(c). $R_{\eta_{s}}$ shows a minimum in the limit $T ->0$ limit and then increases with increase in temperature, $T$. It is always larger than $1$ for all values of $T$. This indicates that fluctuation in Stoke's efficiency, $\sigma_{s}=\sqrt{<\eta_{s}^{2}>-<\eta_{s}>^{2}}$ always dominate the mean value, $<\eta_{s}>$ confirming that $\eta_{s}$ is not a self averaging quantity in this parameter regime. The nature of the distribution function, $P(\eta_{s})$ for various temperature values is studied in Fig.7(d). For lower temperature values, i.e., close to the deterministic limit, the distribution shows a single peaked structure with few minor peaks. With further increase in temperature, the probability of moving the particle to the nearest neighbouring potential minima in the forward direction increases, resulting a finite positive contribution towards the net current, $<j>$ and towards the average Stoke's efficiency, $<\eta_{s}>$. For larger temperature, the barriers disappear in both the directions. As a result the particle spreads over in both the direction. This results decrease in the net current, $<j>$ and average Stoke's efficiency, $<\eta_{s}>$. At the same time the distribution becomes narrow and shifts towards left.   

\begin{figure}
%[hbp!]
\begin{center}
\includegraphics [width=3.5in,height=3in]{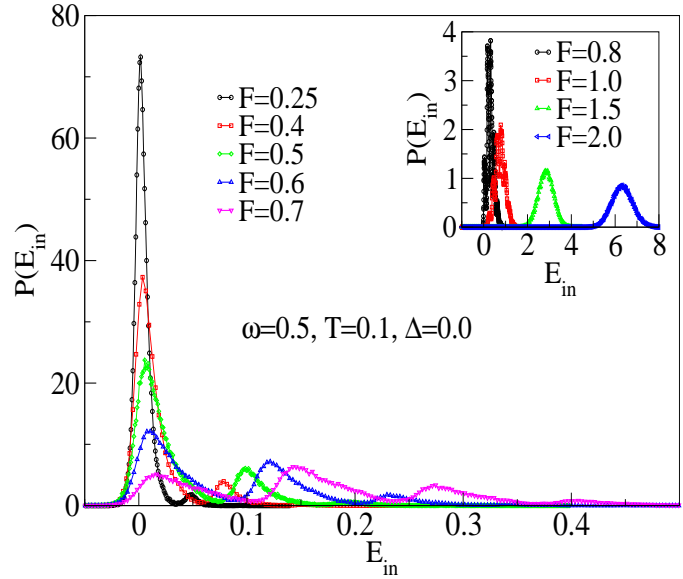}
\caption{The probability distribution of input energy $E_{in}$, $P(E_{in})$ for different $F$ values. The other fixed parameters are $\omega=0.5$, $\Delta=0.0$, $\epsilon=0.4$ and $T=0.1$ respectively.}
\label{sawtooth}
\end{center}
\end{figure}

\begin{figure}
%[hbp!]
\begin{center}
\includegraphics [width=4in,height=3.5in]{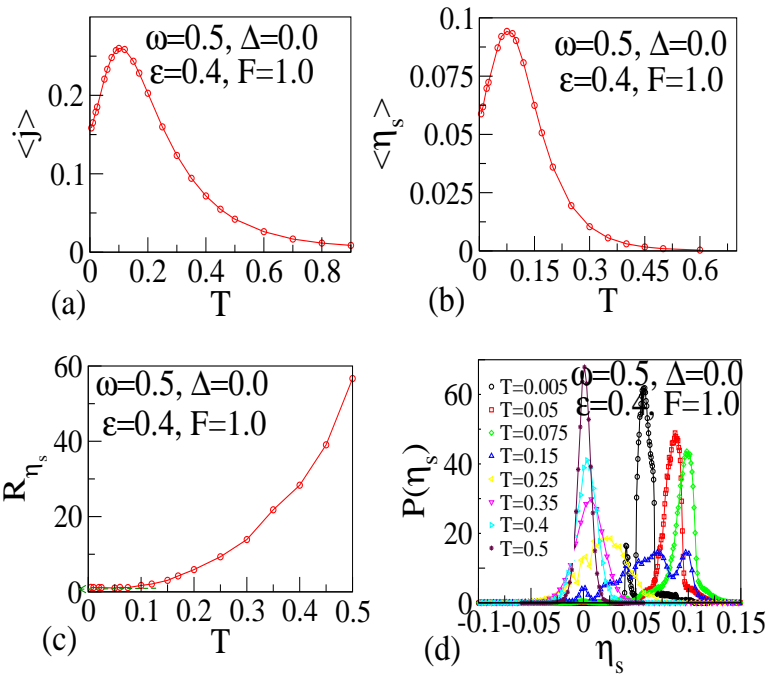}
\caption{$<j>$ as a function of $T$ in (a), $<\eta_{s}>$ as a function of $T$ in (b), relative fluctuation $R_{\eta_{s}}$ as a function of $T$ in (c) and the probability distribution, $P(\eta_{s})$ for different $T$ values in (d). The other fixed parameters are $\omega=0.5$, $\Delta=0.0$, $\epsilon=0.4$ and $F=1$ respectively.}
\label{sawtooth}
\end{center}
\end{figure}

The average current, $<j>$, average Stoke's efficiency, $<\eta_{s}>$, the relative fluctuations in $\eta_{s}$, $R_{\eta_{s}}$ and the distribution function $P(\eta_{s})$ with spatial asymmetry, $\Delta$ is studied in Fig.8(a), (b), (c) and (d) respectively. The average current, $<j>$ (Fig.8(a)) and average Stoke's efficiency, $<\eta_{s}>$ (Fig.8(b)) increase with increase in spatial asymmetry, $\Delta$. This is due to the fact that with increase in asymmetry of the ratchet potential, the barriers in the forward direction starts decreasing and the probability of moving the particle towards the next nearest neighbouring potential minima in the forward direction increases. This results in a finite positive contribution to the net current in the ratchet and hence to the average Stoke's efficiency, $<\eta_{s}>$. Therefore, in this given parameter regime, for a spatial asymmetry ratchet potential, the magnitude of average current is quite larger than that in case of a symmetric ratchet potential and hence the spatial asymmetry in the ratchet helps in enhancing the Stoke's efficiency. However, the relative fluctuations, $R_{\eta_{s}}$ (Fig. 8(c)) shows a decreasing behaviour with increase in spatial asymmetry, $\Delta$, and become less than $1$ for larger spatial asymmetry in the potential. This result reflects that for a larger spatial asymmetry in the ratchet potential, the fluctuations in Stoke's efficiency do not dominate the mean values. As a result $<\eta_{s}>$ behaves as a good physical variable in this parameter regime and a self-averaging physical quantity. The probability distribution, $P(\eta_{s})$ is studied in Fig. 8(d) for various $\Delta$ values. For a symmetric ratchet potential, i.e., for $\Delta=0.0$, the distribution shows a single peaked structure with few minor peaks in it. The inclusion of spatial asymmetry in the potential decreases the barrier height towards the right of the potential. As a result the local velocities in the forward direction increases and the minor peaks in the distribution disappears. Therefore, with increase in $\Delta$ values, the spreading in the distribution decreases at the same time the distribution with mean value shifts towards right indicating that the $<\eta_{s}>$ increases with $\Delta$.

\begin{figure}
%[hbp!]
\begin{center}
\includegraphics [width=4in,height=3.5in]{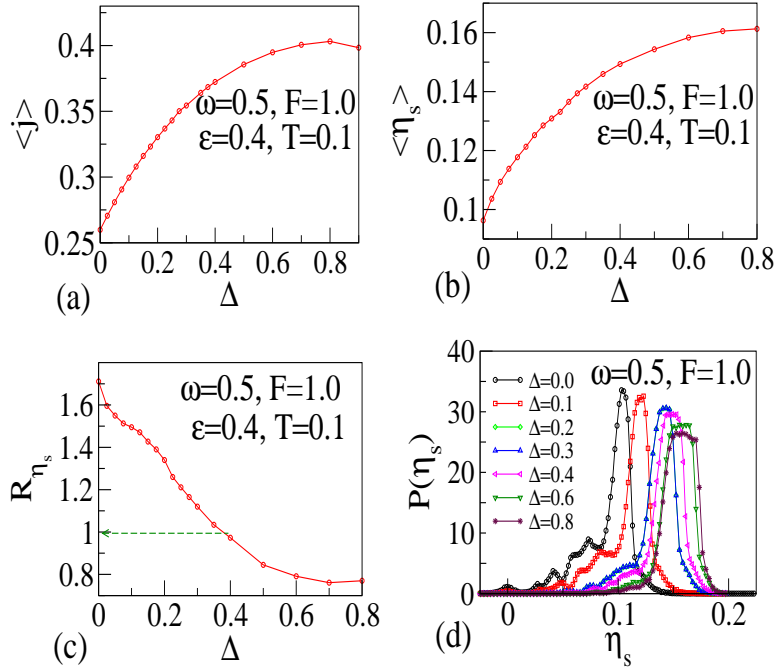}
\caption{$<j>$ as a function of $\Delta$ in (a), $<\eta_{s}>$ as a function of $\Delta$ in (b), relative fluctuation $R_{\eta_{s}}$ as a function of $\Delta$ in (c) and the probability distribution, $P(\eta_{s})$ for different $\Delta$ values in (d). The other fixed parameters are $\omega=0.5$, $F=1$, $\epsilon=0.4$ and $T=0.1$ respectively.}
\label{sawtooth}
\end{center}
\end{figure}

Finally we have studied the average current, $<j>$, average Stoke's efficiency, $<\eta_{s}>$, Relative fluctuations, $R_{\eta_{s}}$ and the probability distribution, $P(\eta_{s})$ with the frequency of driving, $\omega$ in Fig.9(a), (b), (c) and (d) respectively. The average current, $<j>$ (Fig.9(a)) and the average Stoke's efficiency, $<\eta_{s}>$ (Fig.9(b)) shows a decreasing behaviour with frequency of drive, $\omega$. This is because with increase in frequency of drive, the forcing amplitude varies so fast that, the particle is not able to take the advantage of force in either of the direction and the movement of the particle towards the nearest neighbouring potential minima is suppressed. This decreases the net current as well as the average Stoke's efficiency. For very high frequency of driving, the forcing amplitude varies so fast that the particle doesn't feel the influence of forcing. The particle as if stays at one position only and hence the net current, $<j>$ and the average Stoke's efficiency approaches to zero. However, the relative fluctuations, $R_{\eta_{s}}$ (Fig.9(c)) increases with increase in the frequency of driving, $\omega$ and is always larger than $1$. The probability distribution function, $P(\eta_{s})$ with frequency of driving, $\omega$ is shown in Fig.9(d). For low frequency of driving, i.e., in the adiabatic limit, the distribution shows a maximum and with increase in frequency, $\omega$, the distribution spreads over towards left as well as the mean value shifts towards left. With further increase in $\omega$, a finite peak developes across zero with few minor peaks towards the right side of the distribution. At sufficiently high value of $\omega$ (i.e., in nonidiabatic regime), we get a sharp peak across zero and the minor peaks in the distribution get suppressed. This clearly confirms that there is no effective transport in the parameter regime with high frequency of driving.

\begin{figure}
%[hbp!]
\begin{center}
\includegraphics [width=4in,height=3.5in]{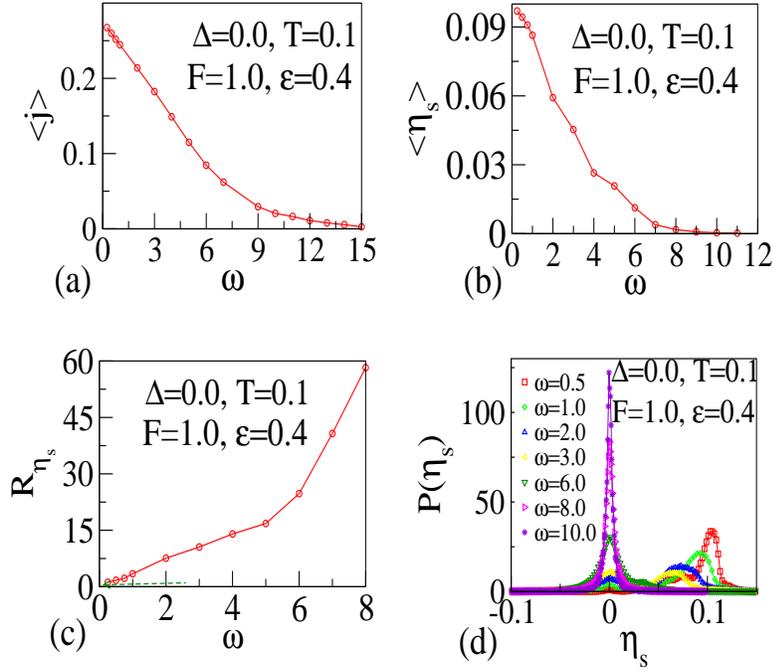}
\caption{$<j>$ as a function of $\omega$ in (a), $<\eta_{s}>$ as a function of $\omega$ in (b), relative fluctuation $R_{\eta_{s}}$ as a function of $\omega$ in (c) and the probability distribution, $P(\eta_{s})$ for different $\omega$ values in (d). The other fixed parameters are $\Delta=0.0$, $F=1$, $\epsilon=0.4$ and $T=0.1$ respectively.}
\label{sawtooth}
\end{center}
\end{figure}

\section{Conclusion}
In conclusion, we have studied the Stoke's efficiency and it's stochastic properties in the case of a temporal asymmetric rocked ratchet potential using overdamped Langevin dynamics simulation. Our main results indicate that over larger parameter space, stochastic Stoke's efficiency behaves as a non-self averaging physical quantity. The mean current and stochastic efficiency monotonically decrease as we go from adiabatic to the non-adiabatic regime. All the above mentioned quantities have been studied for different values of forcing amplitude, temperature, asymmetry parameter and driving frequency. We are presently investigating the nature of stochastic thermodynamic efficiency in these ratchet systems.

\section{Acknowledgment}
MS acknowledges the funding from the INSPIRE faculty award by the Department of Science and Technology, Govt. of India and thanks IOP, Bhubaneswar for the hospitality where a part of the work was done. AMJ thanks DST, India for financial support.


\section{References}
\begin{thebibliography}{0}

\bibitem{Julicher1} F. Julicher, A. Adjari and J. Prost, Rev. Mod. Phys.~ {\bf 69},~ 1269 (1997).

\bibitem{Astumian} R. D. Astumian, Scientific American,~~{\bf 285},~56 (2001).

\bibitem{Reimann} P. Reimann, Phys.Rep.~~{\bf 361}, 57 (2002) and references therein.


\bibitem{jayan} A. M. Jayannavar, Frontiers in Condensed Matter Phys. vol {\bf 5}, ed J K Bhattacharjee and B K Chakrabarti (India: Allied Publishers) p 215[cond-mat/0107079] (2005).

\bibitem{Gammaitoni} L. Gammaitoni, P. Hanggi, P. Jung and F. Marchsoni, Rev. Mod. Phys. {\bf 70} 223, (1998).
\bibitem{Hangi} R. D. Astumian and P. Hanggi,~Phys. Today, {\bf 55},~33 (2002).

\bibitem{marches} P. Hanggi and F. Marchesoni, Rev. Mod. Phys. {\bf 81} 387 (2009).
\bibitem{gommers} R. Gommers, S. Denisov, and F. Renzoni, Phys. Rev. Lett. {\bf 96} 240604 (2006).

\bibitem{Ajdari1} A. Ajdari and J. Prost, C. R. Acad. Sci.. Paris II, {\bf 315},  1635 (1992).

\bibitem{Prost1} J. Prost, J. F Chawini, L. Peliti and A. Ajdari, Phys. Rev. Lett., {\bf 72}, 2652 (1994).
\bibitem{Linke} H.Linke, Appl. Phys. A ~{\bf 75}, ~2 (2002).

\bibitem{Astumian1} R. D Astumian, M. Bier, Phys. Rev. Lett. {\bf 72}, 1766 (1994).

\bibitem{linke1} H. Linke, M. Downtown and M. Zuckermann, Chaos. {\bf 15}, 026111 (2005).

\bibitem{parrondo} J. M. R. Parrondo and B. J. De Cisneros, Appl. Phys. {\bf A75}, 179 (2002).

\bibitem{sekimoto} K. Sekimoto, Stochastic Energetics, Lect. Notes Phys. {\bf{799}} (Springer, Berlin Heidelberg, 2010).
\bibitem{sekimoto1} K. Sekimoto, J. Phys. Soc. Jpn. {\bf 66}, 6335 (1997).

\bibitem{kamgawa} F. Takagi and T. Hondou, Phys. Rev. {\bf E 60}, 4954 (1999); K. Sumithra and T. Sintes, Physica A, {\bf 297}, 1 (2001).

\bibitem{Dan} D.Dan, M. C. Mahato and A. M Jayannavar, Int. J.. Mod. Phys. B {\bf 14}, 1585 (2000); Physica A, {\bf 296}, 375 (2001); ; Phys. Rev E {\bf 63}, 56307 (2001), D. Dan and A. M. Jayannavar, Phys. Rev. {\bf E66}, 41106 (2002).

\bibitem{parrando1} J. M. R. Parrando, J. M. Blanco, F. J. CAo and R. Brito, Euro Phys. Lett., {\bf 43}, 248 (1998).

\bibitem{sokolov} I. M. Sokolov, Phys. Rev. {\bf E63}, 021107, (2001); I. M. Sokolov, cond-mat 0207685v1.
 
\bibitem{hernandez}  N. S\'anchez Salas and A. C. Hern\'andez, Phys. Rev. {\bf E68}, 046125 (2003).

\bibitem{jstat} Raishma Krishnan, Soumen Roy and A. M. Jayannavar, J. Stat. Mech P04012 (2005).

\bibitem{mamata1} R. Krishnan, J. Chacko, M. Sahoo, J. Stat. Mech. {\bf{p06017}} (2006). 


\bibitem{kamgawa-parrondo} H. Kamegawa, T. Hondou and F. Takagi, Phys. Rev. Lett. {\bf 80}, 5251 (1998).


\bibitem{munakata}D. Suzuki and T. Munakata, Phys. Rev. E {\bf 68} 021906 (2003).

\bibitem{oster} H. Wang and G. Oster Europhys Lett {\bf 57} 134 (2002).
\bibitem{wang} H. Wang, Appl. Math. Lett {\bf 22}, 79 (2009).

\bibitem{Bustamante} C. Bustamante, J. Liphardt and F. Ritort. Phys. Today {\bf 58}, 43 (2005).

\bibitem{luczka} J. Spiechowicz, P. Hanggi and J. Luczka, Phys. Rev. E {\bf 90}, 032104 (2014).



\bibitem{rana} S. Rana, P. S. Pal, Arnab Saha and A. M. Jayannavar, Phys. Rev E {\bf 90}, 042146 (2014). 

\bibitem{Mannela}R. Mannela, in:J.A. Freud and T. Poschel(Eds), Stochastic Process i Physics, Chemistry and Biology, Lecture Notes in Physics, vol. {\bf 557} Springer-Verlag, Berlin p353 (2000).


\bibitem{sahoo} M. Sahoo, S. Saikia, M. C. Mahato and A. M. Jayannavar, Physica A: Statistical Mechanics and its Applications, 387 {\bf 25}, 6284 (2008).

\bibitem{Jop} P. Jop, A. Petrosyan and S. Cliberto, EuroPhys. Lett. {\bf 81}, 50005 (2008).

\bibitem{sahoo1} M. Sahoo, S. Lahiri and A. M. Jayannavar, J. Phys. A:Math. Theor. {\bf 44} 205001 (2011).


%\end{thebibilography}
\end{thebibliography}
\end{document}